\documentstyle[preprint,prc,aps]{revtex}
\begin{document}
\draft
\def\relR{\mbox{\small \bf R}}
\def\inr{\mbox{\small \bf r}}
\def\opera{\mbox{\large a}}
\def\larg{\mbox\large g}
\def\Dt{\Delta t}
\def\bR{\bar{\relR}}
\def\bt{\bar t}
\def\tn{\tilde n}
\def\Dt{\Delta t}
\def\Rdot{\stackrel{\cdot }{\relR}}
\def\Rddot{\stackrel{\cdot \cdot }{\relR}}
\newcommand{\etal}{{\it et al\/}\ }
\setcounter{page}0
\title{Dynamic  approach to  fusion  of massive nuclei.}
\author{R.V.Jolos$^{1}$and A.K. Nasirov$^{1,2}$}
\address{Joint Institute for Nuclear Research, 141980, Dubna, Russia}
\author{A.I.Muminov$^{2}$}  
\address{Heavy Ion Physics Department, Institute of Nuclear Physics\\
702132 Ulugbek, Tashkent, Uzbekistan}
\date{\today}
\maketitle

\begin{abstract}

 The role of the entrance channel in the fusion-fission reactions
leading to  nearly the  same superheavy compound nucleus
is studied in the framework of  dynamic model.
The calculations are done for  $^{48}$Ca + $^{244}$Pu and
 $^{74,76}$Ge + $^{208}$Pb reactions which could lead to formation
of superheavy element $Z=114$.  It is shown that for these reactions
there is an energy window for the values of the bombarding energy at which
a capture probability is sufficiently large. Together with the restriction
coming from the intrinsic barrier for fusion, it helps to find  an optimal
value of the bombarding energy for a given projectile--target combination.
\end{abstract}


\section{Introduction}

The  cross section for the production of the superheavy elements  depends
on the  choice  of the  projectile-target combination
and the bombarding energy $E_{\rm c.m.}$. The optimal choice
is determined by the requirements to have  a larger fusion  cross section
and  larger survival probability of a compound nucleus relative to fission.
For a given projectile-target combination, a larger value of the bombarding
energy is needed to overcome the  reaction barrier which is determined by the
nucleus-nucleus potential and the dynamic barriers if they exist.
However,  the excitation energy of the compound nucleus  increases with the
bombarding energy. It decreases the  survival probability relative to fission
of a nucleus produced in a reaction and therefore puts a restriction on
the upper value of a bombarding energy. To determine the optimal value of
$E_{\rm c.m.}$ it is necessary to analyse a dependence of a partial fusion
cross section which is proportional  to a capture probability, on a bombarding
energy. To do it, we require  in  dynamic model to describe
the initial stage of a heavy ion collision. Such a model has been developed
in our earlier papers [1,2] and it is the aim of the present
paper to apply this model to calculate the  capture probability. The latter
quantity is determined by the dynamic aspects of the reaction
mechanism and by the  depth of the  pocket in the nucleus-nucleus interaction
potential. We have calculated a nucleus-nucleus interaction potential
using a double--folding procedure with the Migdal's effective forces [3].
As the examples we consider below the following reactions:
$^{48}$Ca + $^{244}$Pu  and $^{74,76}$Ge + $^{208}$Pb.

\section{Basic formalism}
The cross section of production of the evaporation residues ($\sigma_{er}$)
\begin{equation}
\label{evapor}
\sigma_{er}(E)=\sum_{l=0}^{\infty}(2l+1)\sigma_l^{fus}(E,l)W_{sur}(E,l)
\end{equation}
  is determined
by the partial fusion   cross section ($\sigma_l^{fus}(E)$),
\begin{eqnarray}
\label{s_fus}
\sigma_l^{fus}(E)&=&\sigma_l^{capture}(E) P_{CN}(E,l),\\
\sigma_l^{capture}(E)&=&\frac {\lambda^2}{4\pi}{\cal P}_l^{capture}(E),
\end{eqnarray}
where $\lambda$ is a wavelength, $P_{CN}(E,l)$ is a factor taking into 
account a decrease of the fusion probability due to dinuclear system break 
up before fusion, ${\cal P}_l^{capture}(E)$ is the  capture probability
which depends on the  collision dynamics and determines the amount of
partial waves leading to capture. The cross section of production of the
 evaporation residues $\sigma_{er}$ depends  as well as on
the probability ($W_{sur}(E,l)$) that the compound nucleus survives
during the deexcitation  cascade at the bombarding energy $E$.

To calculate  capture probability  ${\cal P}_l^{capture}(E)$  we shall use
a dynamic approach developed in [1,2]. In this approach, the system of
equations is derived to describe the  radial motion of colliding nuclei and
an evolution of their intrinsic states during the  heavy ion collision.
The relative motion coordinate $\relR(t)$ and the velocity $\Rdot(t)$ are
determined by solving the equations of motion
\begin{equation}
\label{maineq}
\mu(\relR(t))\Rddot_k + \sum_j\gamma_{kj}[\relR(t)]\Rdot_j(t)=
-\frac {\partial W(\relR(t))}{\partial \relR_k}
\end{equation}
where 
$$\mu(\relR)=m A_T A_P/(A_T+A_P)+\delta \mu(\relR),$$ 
$\delta \mu(\relR)$ is the  dynamic contribution to the reduced mass
$\mu$, $\gamma _{kj}[\relR(t)]$ is the  friction tensor,
$W(\relR)=V(\relR)+\delta V(\relR)$ is the  nucleus-nucleus interaction
potential and  $\delta V(\relR)$ is
the dynamic contribution to a nucleus-nucleus potential which is due to
the  rearrangement of the densities of the interacting nuclei during
reaction. To calculate $\delta \mu(\relR)$,   $\gamma_{kj}[\relR(t)]$ and
$\delta V(\relR)$, it is necessary to find the occupation numbers of the
single particle states. Since the excitation energy of the interacting nuclei
changes significantly during the course of the collision, it is necessary to
take into account the time dependence of the occupation numbers. An evolution
of the occupation numbers has been defined by a numerical solution
of the von Neumann equation for the single particle density matrix
$\hat{\tn}$ with the Hamiltonian $\hat H$ which takes the following
form in the second quantized representation
\begin{equation}
\label{DNSsec}
\hat  H(\relR(t),\xi)=\sum_{P} \varepsilon_P
 \opera_P^+\opera_P^{} + \sum_{T} \varepsilon_T \opera_T^+\opera_T^{}+
 \sum_{i, i'} V_{ii'}(\relR(t)) \opera_i^+\opera_{i'}^{} + V_{res}.
\end{equation}
Here $\xi$ is the  short notation for the relevant intrinsic variables,
the third term on the r.h.s. of the Eq. (\ref{DNSsec}) can be written
as
\begin{eqnarray}
\label{coupt} && \sum_{i, i'} V_{ii'}(\relR(t)) \opera_i^+\opera_{i'}^{}
=\sum_ {P, P'} \Lambda^{(T)}_{PP'}(\relR(t))\opera_P^+\opera_{P'}^{} +
\sum_{T, T'} \Lambda^{(P)}_{TT'}(\relR (t))\opera_T^+\opera_{T'}^{} + \\
&&\sum_{T,P} \mbox{\large g}_{PT}(\relR (t))(\opera_P^+
\opera_T^{} + {\rm h.c.})\,,
\nonumber
\end{eqnarray}
where  $P\equiv (n_P,j_P,l_P,m_P)$ and $T\equiv (n_T,j_T,l_T,m_T)$ are
the sets of quantum numbers characterizing the single particle states in
the noninteracting  projectile and the target nuclei, respectively.
The last term on the r.h.s. of Eq. (\ref{DNSsec}) represents the
residual interaction. Since an explicit allowance for the residual
interaction is very complicated it is customary to take into account
a two--particle collision integral in the linearized form
($\tau$--approximation) [1,2,4-6]
\begin{eqnarray}
\label{eqoccnum}
i\hbar \frac {\partial \hat{\tn}(t)}{\partial t} =
[\hat {\cal H}(\relR(t)),\hat {\tn}(t)] -
 \frac {i\hbar}{\tau}[\hat {\tn}(t) -  \hat {\tn}^{eq}(\relR(t))]\,,
\end{eqnarray}
where
$\tn^{eq}(\relR(t))$ is the  local quasi-equilibrium
distribution, {\it i.e.} a Fermi distribution with the temperature $T(t)$
corresponding to the  excitation energy at the internuclear distance
$\relR(t)$. All formulae needed to calculate  $\gamma_{kj}[\relR(t)]$
and $\delta V(\relR))$ are given in [2,4,6].

The nuclear part of a nucleus-nucleus potential $V(\relR(t))$ is calculated
using the double-folding procedure between the effective nucleon-nucleon 
forces $f_{eff}[\rho(x)]$ suggested by Migdal [3] and  the densities  of
the interacting nuclei taken in the Woods--Saxon form
\begin{equation}
\rho^{(0)}_K({\bf r},\relR_K(t),\theta_K,\beta_2^{(K)})=
\biggl[1+\exp\Bigl(\frac{|{\bf r}-\relR_K(t)|-R_{0K}(1+
\beta_2^{(K)}Y_{20}(\theta_K))}{a}\Bigr)\biggr]^{-1},
\end{equation}
 where $\relR_K$  are the center of mass coordinates and $R_{0K}$ are the 
half density  radii of interacting nuclei $K=1,2$; $\beta_2{(K)}$ are the
quadrupole deformation parameters determined by the $B(E2)$ to  the 
first-excited $2^+$ state (its value is taken from [7]) and $\theta_K$ are 
the axial symmetry axes orientations relative to $\relR(t)$. Thus, we have
a possibility to consider fusion at different mutual orientations of
the interacting nuclei.

The competition between complete fusion and quasifission of a dinuclear
system formed after capture and  its further evolution are described
using the method developed in [8].
This method is based on the assumption that dinuclear system  formed
in the collision of two nuclei evolve to fusion by increasing
its mass asymmetry. It means that the mass asymmetry degree of freedom
$\eta=(A_T-A_P)/(A_T+A_P)$ is the main dynamic variable.
The internuclear distance $R(t)$ takes the value corresponding to the
location of the minimum of the nucleus-nucleus interaction potential for
every given value of $\eta$. The  evolution of the system along mass 
asymmetry degree of freedom is described by a driving potential $U(Z,l)$ 
which is calculated as
\begin{equation}
\label{drivp}
U(Z,l)=B_1+B_2+U_{12}(R_m)-B_0.
\end{equation}
Here, $B_1$ and $B_2$ are the binding energies of the
nuclei in a dinuclear system, $U_{12}(R_m)$ is the value of the
nucleus--nucleus interaction potential at the minimum, $B_0$ is the binding
energy of the compound nucleus (the binding energies $B_i$ are obtained from
[9] and from [10]  particularly for the superheavy elements). Therefore, 
a dinuclear system to be fused should overcome the intrinsic barrier 
($B^*_{fus}$) which is determined by the difference between the values
of a driving potential located at the Businaro--Gallone
point $(\eta=\eta_{BG})$ and the initial point corresponded to reaction under 
consideration. For the reactions considered below, the initial value of 
$\eta$ is smaller  than  $\eta_{BG}$.  The quasifission, which is in  
competition with the  fusion is considered as a motion in the nucleus-nucleus
interaction potential $W(R)$. Thus, for quasifission, it is necessary to
overcome a barrier of $W(R)$. The competition of fusion and quasifission
is taken into account by the factor $P_{CN}(E,l)$, which is
calculated using the following relation  derived from  the
statistical model arguments
\begin{equation}
\label{Pcn}
P_{CN}=\frac{\rho(E^* - B^*_{fus})}{\rho(E^* - B^*_{fus})
+ \rho(E^* - B^*_{qf})}.
\end{equation}
Here  $\rho(E^* - B^*_K)$ is the  level density
\begin{eqnarray}
\label{levden}
\rho(E^* - B^*_{K})&=&\frac{g(\epsilon_F)K_{rot}}
{2\sqrt{g_1(\epsilon_F)g_2(\epsilon_F)}}
\frac{\exp\left[2\pi\sqrt{g(\epsilon_F)(E^* -B^*)/6}\right]}
{\left[\frac 32g(\epsilon_F)(E^* -B^*)\right]^{\frac14}
(E^* -B^*_K)\sqrt{48}}.
\end{eqnarray}
In Eq. (\ref{Pcn}), $B^*_{fus}$ is the barrier of the driving potential
$U(Z,l)$, which should be overcome on the way from the initial value
of $\eta$ to $\eta=1$. The $B^*_{qf}$ is the barrier of the nucleus-nucleus
interaction potential which should be overcome if dinuclear system  decays
in two fragments,
$E^*$ is an excitation energy of the compound nucleus which is  equal
to difference between $E_{\rm c.m.}$ and the minimum of nucleus--nucleus
potential ($E^*=E{\rm c.m.}-U(R_m)$),  $g_{1,2}(\epsilon_F)$
are the single particle level densities of the fragments of the dinuclear
system  and $g(\epsilon)$ is their sum, $K_{rot}$ is a factor taking into
account rotation of a dinuclear system
\begin{equation}
\label{Krot}
K_{rot}=\frac{\sqrt{6(E^* -B^*)/g(\epsilon_F)}}{\pi}J_{\perp},
\end{equation}
where $J_{\perp}$ is the rigid body moment of inertia for rotation around the
axis perpendicular to the line connecting the centers of fragments.

\section{Results and discussion}

We  consider below the following reactions which are discussed
now as possible ways to search for superheavy element with $Z=114$.
They are  $^{48}$Ca + $^{244}$Pu (I) and $^{74,76}$Ge + $^{208}$Pb (II,III).

Basing on the  dynamic model developed in [1] (which is described concisely
in the preceding section) we have calculated the capture cross section
$\sigma_l^{capture}(E)$ for the reactions under consideration.
The results are shown in Fig. 1. It is seen that, for these reactions there
is an  energy window for the values
of the bombarding energy at which  a capture cross section is  large enough
to have a physical interest. The lower limit for the bombarding energy
$(E_{min})$ is defined by a total nucleus-nucleus interaction potential
$W(R)=V(R)+\delta V(R)$. Note that $E_{min}$ is somewhat larger than the value
of the entrance Coulomb barrier,  because of the  kinetic energy loss due to
friction. So, $E_{min}$ is determined by a dynamic calculation.
The upper limit $(E_{max})$ comes from an incomplete dissipation of the
relative kinetic energy. Thus, the values of $E_{min}$  and $E_{max}$  are
determined by the  depth of the  pocket in the potential $W(R)$ (Fig. 2) and
by dissipative forces. If a bombarding energy is larger than $E_{max}$ the
dissipative forces could not provide a complete dissipation of the relative
kinetic energy and dinuclear system decays into two fragments instead of
being fused. As it is seen from Fig. 1, reaction with the lighter projectile
(I) has a larger value of the capture cross section than other two reactions
(II) and (III). The reason is that for $^{48}$Ca + $^{244}$Pu reaction the
pocket of the nucleus--nucleus interaction potential is deeper and wider than
for  $^{74,76}$Ge + $^{208}$Pb (see Fig. 2). The potentials presented
in Fig. 2 are calculated taking into account a deformation of the interacting
nuclei assuming the tip-tip orientation of the colliding projectile and target
nuclei. For other orientations of the colliding nuclei the potential
is more flat and the  depth of the  pocket is smaller. Moreover, in these
cases an entrance barrier and the minimum of the  pocket of $W(R)$
have larger absolute energies than in the case of the tip-tip orientation.
Therefore, an excitation energy of a compound nucleus will be larger than
in the last case. An excess of the excitation energy will decrease the
survival probability of the evaporation residues. Thus, in the fusion of
massive nuclei their mutual orientation strongly influences not only  the
capture cross section but also the probability that the compound
nucleus survives during deexcitation.

The existence of the window for the bombarding energy has a crucial
influence on the fusion process. From one side a larger bombarding energy
will be needed to overcome an intrinsic barrier ($B_{fus}^*$) to form a
compound nucleus. From other side an increase of the bombarding energy
decreases the capture probability  starting from some values of the
bombarding energy because the  friction force is not strong enough to provide
a complete dissipation of the kinetic energy.

To analyse a fusion process further, we  need in a dynamic model which
describes an evolution of a dinuclear system to compound nucleus. Below,
we shall use a model developed in [8]. According to this model a dinuclear
system evolves to compound nucleus by increasing its mass asymmetry. It means 
that driving potential (\ref{drivp}) plays the main role in the fusion 
dynamics and a dinuclear system should overcome the Businaro--Gallone point 
to be fused. The driving potentials for the reactions which we analyse are
presented in Figs. 3-5. The values of the barriers which should be overcome
to get a compound nucleus $(B^*_{fus})$ depend on the compound system and
the reaction choice which determines the initial value of the mass asymmetry.
These are equal to 6 MeV for $^{48}$Ca + $^{244}$Pu (Fig. 3) and 28 MeV for
$^{74,76}$Ge + $^{208}$Pb (Figs. 4 and 5).  To overcome the barrier, a
dinuclear system should have the corresponding excitation energy.
However, the possible values of the excitation energy which are defined
by the amount of a dissipated energy are restricted by the framework
of the energy window for bombarding energies leading to capture.
The possible values of the excitation energy can be estimated and the results
are shown in Fig. 6. For   $^{48}$Ca + $^{244}$Pu reaction  the excitation
energy can take the values from 19 MeV up  to 41 MeV
which are larger than the barrier $B_{fus}^*$ of the driving potential.
In the case of $^{74,76}$Ge + $^{208}$Pb reactions, the excitation energy
$E^*$ takes the values between 6 MeV and 16 MeV. This value is lower than
the value of $B_{fus}^*$=28 MeV for these reactions but it is larger than
the quasifission barrier which is about $(3-5)$ MeV (Fig. 2).
 An increase of the beam energy in order to obtain an adequate
excitation energy does not help because dinuclear system can not be
formed. The corresponding value of the beam energy will exceed $E_{max}$.
Thus, according to our calculations of a capture cross section and the
model of fusion suggested in [8], the compound nucleus can not be formed
with a measurable cross section in the $^{74,76}$Ge + $^{208}$Pb reactions.
However, it is not excluded that a dinuclear system can prefer the trajectory
in the $R-\eta$ plane for fusion different from that suggested in [8] or other
mechanism of the compound nucleus formation like cluster transfer [11]
might play an important role.

The other question  concerns the  probability that the excited compound
nucleus formed in a fusion process survives during deexcitation. An increase
of an excitation energy decreases the influence of the shell effects
on stability of a compound nucleus and decrease the fusion probability.
However, this question is not analysed in the present paper.

\section{Conclusion.}

We have analysed the partial fusion cross sections for the reactions with
massive nuclei leading to compound nucleus with $Z=114$:
$^{48}$Ca + $^{244}$Pu  and $^{74,76}$Ge + $^{208}$Pb.
The main attention is paid to the calculations of the capture probability,
which is a characteristic feature of an initial stage of the collision.
It is shown that for the considered reactions, there is an energy window
for the bombarding energy at which the  capture cross section is large enough
to have a physical interest.
This result puts a strong limitations on the  choice of the bombarding energy
for a given reaction. However, from other side, the excitation energy should
be large enough to overcome an intrinsic barrier for the fusion [8].
Thus, both restrictions can be used to obtain an optimal choice of the
projectile-target combination and of the bombarding energy.\\

\acknowledgements

We are grateful to Prof. V.V. Volkov for the fruitful discussions.
This work was supported by the Russian Fund for the Fundamental
Research Grant 97-02-16030.

\newpage
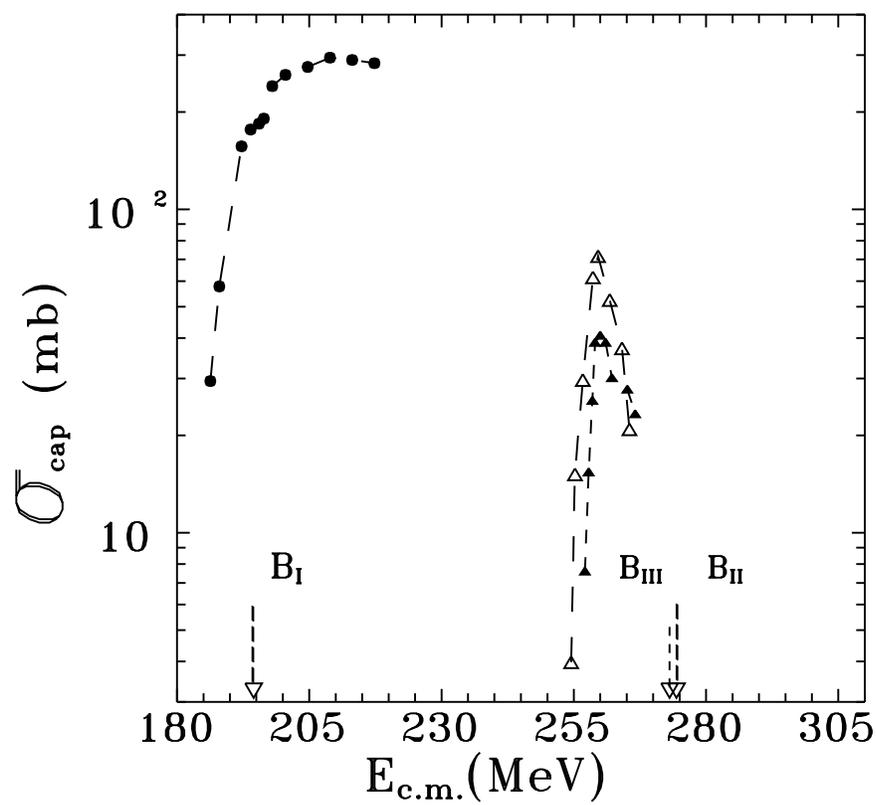
\begin{figure}
\caption{The calculated capture cross section as a function of the
 beam energy for the $^{48}$Ca + $^{244}$Pu (I)
 (full circles), $^{74}$Ge + $^{208}$Pb (II) (full triangles), and
 $^{76}$Ge + $^{208}$Pb (III) (open triangles) reactions;
B$_{i}$ is the Bass barrier  for the reaction (i), i=I, II, and III.}
\label{Fig. 1}
\end{figure}

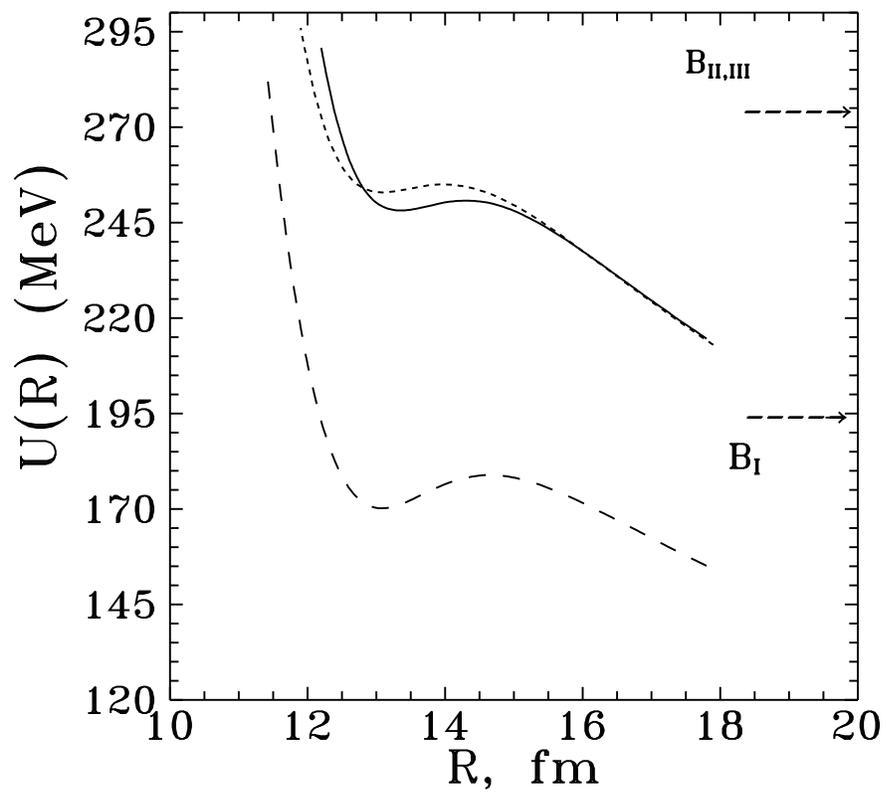
\begin{figure}
\caption{The nucleus-nucleus interaction potential calculated for
$^{76}$Ge + $^{208}$Pb (solid curve), $^{74}$Ge + $^{208}$Pb (dotted curve)
and  $^{48}$Ca + $^{244}$Pu (dashed curve) reactions;
 B$_{i}$ is the Bass barrier
 for the reaction (i), i=I, II, and III.}
\label{Fig. 2}
\end{figure}

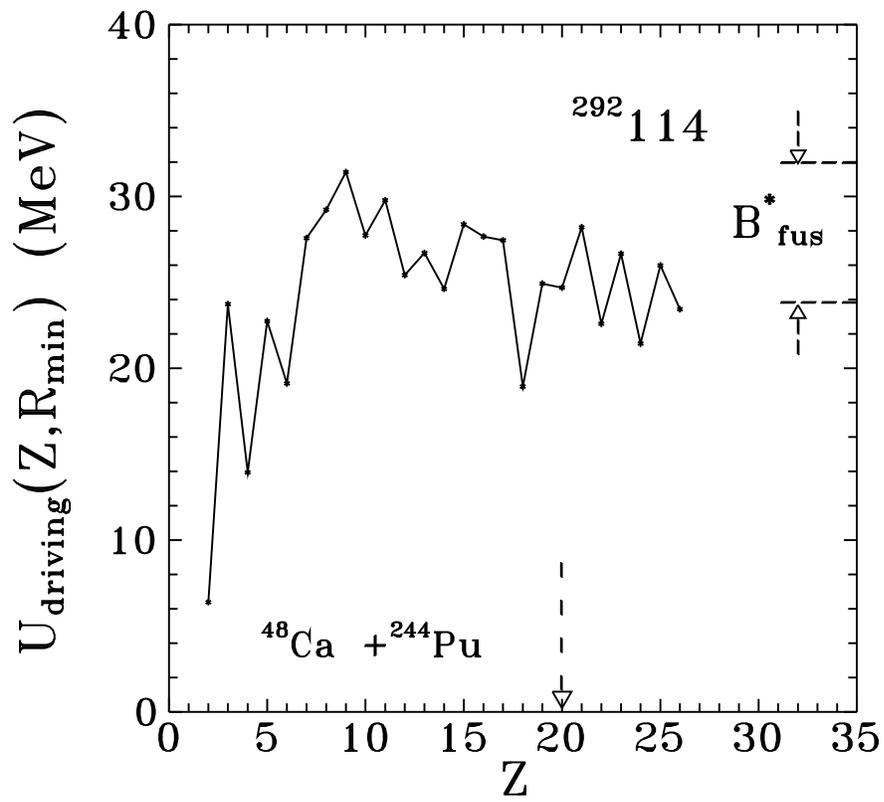
\begin{figure}
\caption{The driving potential for the superheavy element $^{292}$114.
The arrow indicates an initial charge asymmetry which corresponds to 
the $^{48}$Ca + $^{244}$Pu reaction.}
\label{Fig. 3}
\end{figure}

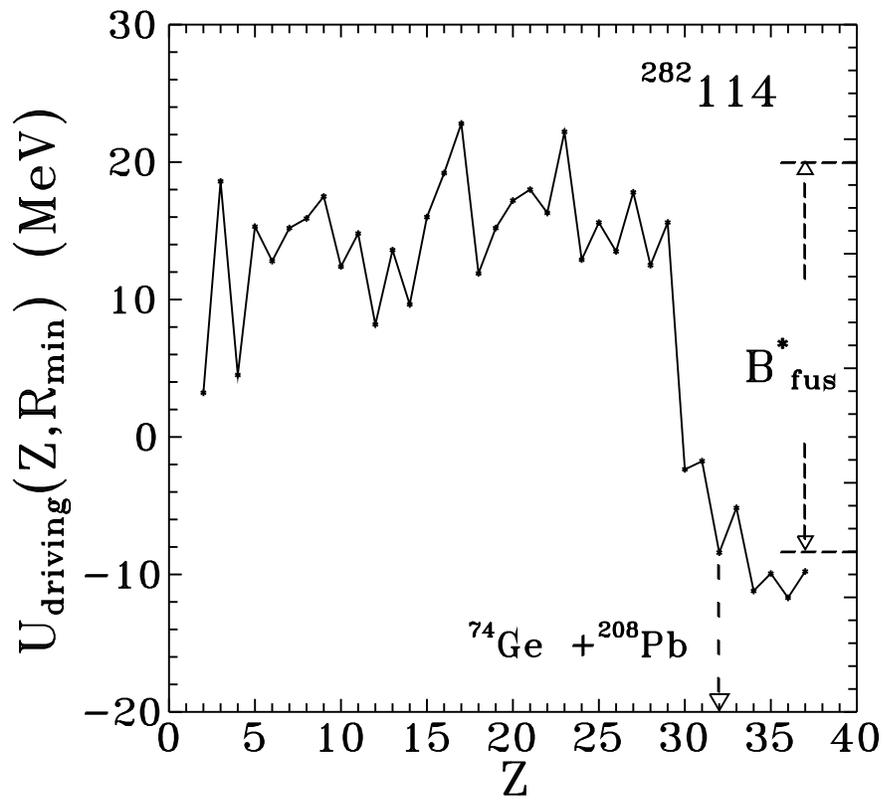
\begin{figure}
\caption{The driving potential for the superheavy element $^{282}$114. 
The arrow indicates an initial charge asymmetry which corresponds to 
$^{74}$Ge + $^{208}$Pb reaction.}
\label{Fig. 4}
\end{figure}

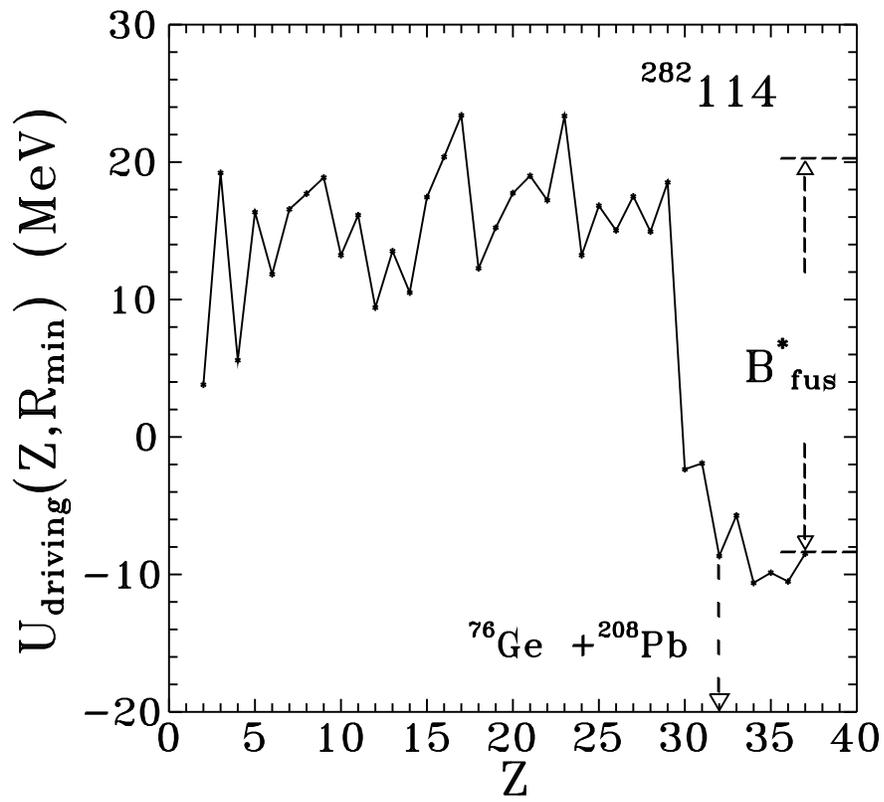
\begin{figure}
\caption{The driving potential for the superheavy element $^{284}$114. 
The arrow indicates an initial charge asymmetry which corresponds to 
$^{76}$Ge + $^{208}$Pb reaction.}
\label{Fig. 5}
\end{figure}

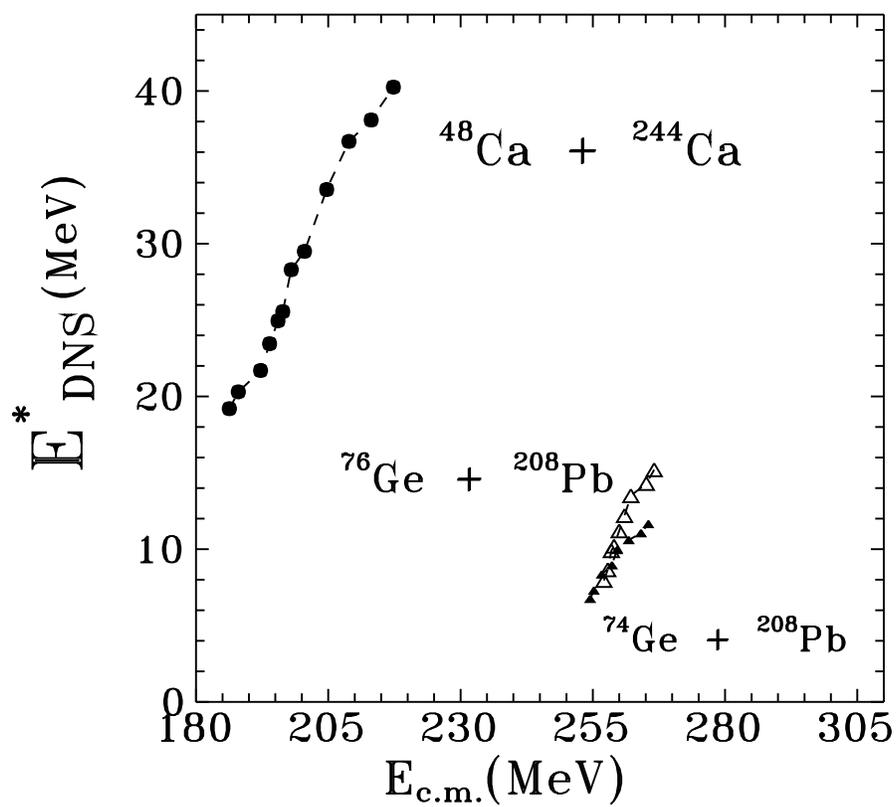
\begin{figure}
\caption{The excitation energy of a dinuclear system formed
after capture of nuclei in reactions:  $^{48}$Ca + $^{244}$Pu (I)
 (full circles), $^{74}$Ge + $^{208}$Pb (II) (full triangles), and
 $^{76}$Ge + $^{208}$Pb (III) (open triangles) as a function of the 
 beam energy in the center of mass system.}
\label{Fig. 6}
\end{figure}

\end{document}